\documentclass[12pt]{iopart}

\usepackage{iopams}
\usepackage{calc}

\usepackage[utf8]{inputenc}
\usepackage[T1]{fontenc}
\usepackage[magyar,english]{babel}
\usepackage{graphicx}
\eqnobysec

\begin{document}

\title[]{Retarded cosmological gravity and Mach's principle in flat FRW universes  }
\author{Bal\'azs  Vet\H o}

\address{Department of Physics of Roland E\"otv\"os University,
P\'azm\'any P\'eter s\'et\'any 1/A. Budapest, H{--}1117,  Hungary}
\ead{veto.balazs@metal.elte.hu}

\begin{abstract}
The  retarded  gravitation produced by  the matter and energy content of the observable   universe is formulated  and shown how this cosmological gravity  gives rise to inertial forces in accelerated frames of reference. The model is developed for spatially flat  Friedman-Robertson-Walker (FRW)  universes.

The retarded potential of cosmological gravity  is determined first   in    fundamental  frames of reference,  in Einstein-de Sitter (ES)
 and  in  $\Lambda$CDM   models. 
Despite  the  expansion, the observable universe gives rise to a static, retarded  gravitational  potential of $\mathit{\Phi}_E = -\, c^2$  in ES universe, while gravity is weakening  during the expansion  in $\Lambda$CDM model.

In the locally  Minkowski spacetime occurring in a small domain of the universe, a covariant
 Lagrangian of a test body moving in the cosmological gravitational field is derived. 
 Inertial forces   are determined from a test body's Euler-Lagrange equation, in Newtonian approximation. 
  The model calculation indicates, that  
the  inertial forces     meet exactly the ones observed  in ES universe and Mach's principle holds there. In the present-day standard, $\Lambda $CDM model 
    inertial forces are  weaker  by 10 \% than  expected.   

\end{abstract}

\submitto{Class. Quantum Grav.}

\maketitle

\section{Introduction}
\subsection{ The cosmological origin of inertial forces}

The experimental evidence that   inertial forces   occur   in frames of reference accelerated  relative to 
  inertial frames was explained by the effect  of the absolute space   in the Newtonian mechanics. The idea that inertial forces are real and not fictitious ones and  originate from the mass of the universe was first formulated by Mach [1]. The general theory of relativity (GR) helped us to  understand the mechanism  how the  gravitation  of the universe is able generate such forces. However, Einstein [2] himself was never entirely certain whether GR incorporates Mach's principle. Referring to the  publications edited by Barbour and Pfister [3],  the validity of Mach's conjecture concerning on the origin of inertial forces was  contested still by the end of the last century.  

Some investigations hint that inertial forces are generated by the cosmic matter. At the beginning, Thirring [4] suspected inertial like forces inside a rotating mass shell using GR. Several decades later, Brill and Cohen [5]  showed that GR predicts exactly the experienced centrifugal and Coriolis forces to  occur in the centre of  a rotating mass shell (a simple model of universe), when the Schwarzschild radius of the shell  is equal to its radius.
 The phenomenon is called perfect or exact dragging. If the observable universe  matched the condition of perfect dragging, then  it would exert the expected inertial forces in frames of reference accelerating or rotating relative to its mass distribution and Mach's principle would be valid. 

Mach's principle holds in FRW  universes, in which the metric arises from material source alone.  Raine [6]  based this assertion on the analysis of Riemann curvature and  of the  stress tensor.  He stated furthermore, that Mach's principle does not hold in empty universe and in models in which  anisotropic expansion or rotation is present.

Model calculations were carried out  by Schmid [7], who investigated  frame dragging in  flat, $k=0$ FRW   universes  and demonstrated their Machian nature. He found perfect dragging  considering cosmological gravitomagnetism and  linear vorticity  perturbation in FRW spacetime.  More recently,  Schmid [8] presented that Mach's principle holds    also in   FRW universes $k=\pm 1$, even in cases when the cosmological constant is not zero. 
 Schmid's  results   indicate that GR incorporates  Mach's principle and perfect dragging is present in all type of  the FRW models investigated. However, the concept of the "instantaneous action-at-a-time" used in the cosmological gravity might  conflict with GR, hurt the finite speed of propagation of gravity.
To ensure  finite gravitation of an infinite universe, Yukawa potential is used
     in Schmid's model. The gravitation of masses beyond the H-dot radius are ruled out by the exponential cut off of the potential.

One can obtain a finite    effect of cosmological gravitation under more realistic conditions, by  keeping the  propagation speed  of gravity finite.  This approach yields automatically  a finite gravitation generated by  the finite,  observable part of the universe and Yukawa potential is spared.  In addition to the   above  investigations,  some contributions are presented about the origin of inertial forces and  their relation to the  retarded cosmological gravitation in flat FRW universes.

\subsection{The goal and the physical scope of present model}

Present model is intended to develop a theory, which explains how the gravity of the observable part of the universe gives rise to inertial forces in accelerated frames of reference,  in spatially flat, FRW spacetime. To draw up the physical approach of  cosmological gravity, several postulates have to be made:
\begin{enumerate}
\item  The observable universe has a locally Minkowski, asymptotically FRW metric, as observers really perceive the spacetime in their laboratories and at distant galaxies, respectively. 
\item  All the elements of the entire matter and energy content of the observable universe gives rise to elemental gravitation.  The sum of the elemental gravitations  of the observable universe is the cosmological gravity.
\item Gravitation is assumed to be transmitted by a flow of gravitons, propagating at the  speed of light. 
\item An observer experiences  the gravitation of those  elements of the universe from where the emitted  gravitons got to him in the moment of the observation. 
\item The gravitation  of the different  elements of the observable universe, which is perceived by an observer   at present cosmic time, was emitted sometime  in the  past. The moment of emission, $t_e$ and of  detection, $t_0$  is measured in cosmic time.  The time difference of $t_0-t_e$, called retardation,
  depends on the coordinate distance between the location of detection and   emission and on  the rate of the expansion, i.e.  on the model universe. 
 Because of the strong similarity of propagation of  photons and gravitons, the effective cosmographic distance of gravitation is taken to be identical with the photon flux distance, see Visser and Catto\"en [9]. 
\end{enumerate}

\noindent In our scenario,  the  retarded gravity of the observable  universe is determined, first  in a fundamental frame of reference,   as the  frame of reference at rest in the mass distribution of the universe was named by Peacock [10]. 
The retarded gravitational potential of a mass element, perceived by an observer at present cosmic time is written as usual in Newtonian dynamics,
\begin {equation}
\mathrm{ d} \mathit{\Phi}(t_0) =- G\frac{{\rm d} m}{d_F(t_0,t_e)}.
\end {equation} 
Here $d_F(t_0,t_e)$ denotes the  photon flux  distance, i.e. the effective cosmographic distance of gravity  between the mass element  and the observer, which involves the  effect of retardation by   representing the model universe and the moment of the emission and detection. 
The retarded gravitational potential is calculated   by
  summing up all the elemental gravitational potentials, which  are perceived at present time $t_0$.    The observable cosmological gravitation is determined  in ES 
 and in the current standard,  $\Lambda$CDM
 universes with parameters of
 $\Omega_{\Lambda 0} =0.728$ and $\Omega_{m0}=0.272$.

Subsequently,   a covariant description of cosmological gravity is given.
   Despite  the presence of  the strong cosmological gravity,   special relativity (SR) to holds in  locally Minkowski spacetime.
Using SR and the Lorentz transformation between instantaneous inertial frames, the relativistic Lagrangian   of a  test particle moving in the gravitational potential field of the universe is determined in accelerated frames of reference.  
  At last,   the inertial forces generated by cosmological gravity are investigated  assuming non-relativistic velocities, i.e.  in Newtonian approximation of the Euler-Lagrange equations.

\subsection{Basics of Einstein-de Sitter and ${\mathit \Lambda}$CDM  universes}

  Notations and basic equations are summarized below, which are needed to  determine the gravitational potential in the ES and $\Lambda$CDM universes.     
In  a spatially flat   FRW universe,  the metric is written simple, 

\begin{equation}
\mathrm{d}s^2 = c^2\mathrm{d}t^2-a^2(t)(\mathrm{d}R^2 + R^2\mathrm{d}\vartheta^2+ R^2\sin ^2\vartheta \mathrm{d}\varphi^2).
\end{equation}
 Here $t$ denotes the cosmic time, $R,\;  \vartheta, \; \varphi$ represent dimensionless, co-moving, spherical coordinates centered at a fundamental observer.   The scale factor, $a(t)$  describes the rate of  expansion of the universe and   has a dimension of length. The subscript $0$ refers to the present time value for any quantity. 
 Therefore $a_0\equiv a(t_0)$, denotes the present value of the scale factor, which   is the length parameter itself.

The   model  universes investigated are homogeneously filled by  pressure free matter. The matter density of $\rho_m$ decreases in the course of the expansion.
In the $\Lambda$CDM model one supposes  a vacuum or dark energy to be present with a contribution to the density of $\rho_\Lambda$. The density of vacuum energy is constant, so
the total density of the universe at a cosmic time of $t$  may be written, as
\begin{equation}
 \rho(t) = \rho_{m}(t)+\rho_\Lambda = \rho_{m0}a^3_0/a^3(t) + \rho_\Lambda .
\end{equation}
The cosmological constant is defined by $\Lambda=8\pi G\rho_\Lambda$, where $G$ is for the gravitational constant.
The Friedman equation  yields connection between scale factor and density of the universe. In the case of $k=0$, it takes the form of 
\begin {equation} 
\Bigl(\frac{\dot {a}}{a}\Bigr)^2 =\frac{8\pi G[ \rho_m(t)+\rho_\Lambda]}{3}, 
\end {equation}
 as   deduced by Berry [11]. Solving the Friedman equation, one obtains the scale factor as a function of cosmic time, 

\begin{equation}
a(t) = a_0 \sqrt[3]{6\pi G\rho_{m0}}\; t^{2/3}, \qquad \mathrm {if}\qquad k=0,\qquad \Lambda =0,\qquad \mathrm {and}
\end{equation}

\begin{equation}
a(t)=a_0\sqrt[3]{\frac{\rho_{m0}}{\rho_\Lambda}}\;\sinh^{2/3}\Big(\frac{\sqrt{3\Lambda}}{2}t\Big), \qquad \mathrm {if}\qquad k=0,\qquad \Lambda >0.
\end{equation}

The scale factor is coupled to the Hubble parameter and to the cosmological redshift by the connections
$ H(t) ={ \dot{a}(t)}/{a(t)} $   and $1+z =   {a_0}/{a(t)}$.
The photon flux distance, used in equation (1.1),  is determined by the co-moving coordinate of the emitting   mass element of $R_e$ and by the scale factor.
  In spatially flat FRW universes, 
\begin{equation}
d_F(t_0, t_e)={a_0^{3/2\,}}{a^{-1/2}(t_e) R_e},
\end{equation}
 referring to Visser and Catto\"en [9].

\section{Retarded cosmological gravity in fundamental frames of reference }
\subsection{The retarded gravitational potential of the observable universe}

To obtain the retarded potential of cosmological gravity, let us sum up  the retarded elemental  potentials  of the observable universe, given in equation (1.1),

\begin {equation}
\mathit\Phi_U(t_0, 0)=-G \int_{V_{obs}(0)}\frac{\rho(t_e)}{d'_F(t_0,t_e)}\textrm{d}V'_e.
\end{equation}
$ \mathit\Phi_U(t_0, 0)$ denotes the potential of the observable universe  in the origin of the coordinate system and $V_{obs}(0)$ is the volume of the observable  universe around the origin. In agreement with the cosmological principle, the potential of the cosmological  gravity does not depend on position, i.e. $\mathit\Phi_U(t_0)= \mathit\Phi_U(t_0, 0)= \mathit\Phi_U(t_0,{ \bf R})$,  because every location of ${\bf R}$ possesses its own observable universe around itself.  Only the respective, observable universe determines the physical conditions in each point of the universe.
 Denoting the objects horizon by $R_{max}$ and let  $R\ll R_{max}$, the above attribute of the potential can be expressed, as 
\begin {equation}\hskip -0.5 cm
\mathit\Phi_U(t_0, {\bf R})=-G \int_{V_{obs}(0)}\frac{\rho(t_e)}{d'_F(t_e)}\textrm{d}V'_e =-G \int_{V_{obs}({\bf R})}\frac{\rho(t_e)}{\vert{\bf d'}_F(t_e)-a_0{\bf R}\vert}\textrm{d}V'_e.
\end{equation}

Because of the spherical symmetry, $\textrm{d}V'_e = 4\pi a^3(t_e)R_e'^2\textrm{d}R'.$
The $R'_e$ denotes radial co-moving coordinate of the location of emission. 
Substituting  $\textrm{d}V'_e$  and $d'_F$ from  equation (1.7)  into equation (2.1), the  gravitational potential takes the form of

\begin{equation}
{\mathit \Phi}_U(t_0)  =- 4\pi G\int_0^{R_{max}} \rho(t_e)  a^{3.5}(t_e) a_0^{-1.5}R'_e\textrm{d}R'.
\end{equation}

Let us consider a beam of light  which  gets  just  at $t_0$ to the observer and was  emitted at  a co-moving coordinate $R_e $,  into the direction to the observer at  a time $t_e$.  The kinematics of the FRW models yields a definite connection between $R_e$ and $t_e$.
Light and gravity travel along  geodesic curves, locally
straight lines, satisfying $ \textrm{d}s = 0$.
 In  cases when light travels along a radial line, the coordinates $\vartheta$ and $\varphi$ are constant and from equation (1.2) we find
\begin{equation}
c\textrm{d}t = a(t)\textrm{d}R.
\end{equation}
  Integrating equation (2.4), the relation, called the function of retardation is obtained,
\begin{equation}
R_e(t_0,t_e)= c\int _{t_e}^{t_0} \frac {\textrm{d}t}{a(t)} .
\end{equation}
  Equation (2.4) and (2.5) make it possible to express  $\textrm{d}R'$ and $R'_e$ by $\textrm{d}t$ and $t_e$ in equation (2.3) and

\begin{equation}
{\mathit \Phi}_U(t_0)  =- 4\pi Ga_0^{-1.5}c^2\int_0^{t_0} \rho(t_e)  a^{2.5}(t_e)\int_{t_e}^{t_0}\frac{\textrm{d} t}{a(t)} \textrm{d}{t_e}.
\end{equation}
The general formula  in equation (2.6)  enables to calculate the retarded  potential of cosmological gravity in flat FRW universes. When calculating the potential,  density and scale factor have to be taken from the respective model universe.

Let us separate  the gravitational  potentials originate from  matter and from vacuum or dark energy, $\mathit{\Phi}_U=\mathit{\Phi}_m +\mathit{\Phi}_\Lambda$. Replacing density from equation (1.3) into equation (2.6),
 
\begin{equation}
{\mathit \Phi}_m(t_0)  =- 4\pi G c^2 \rho_{m0}   a^{1.5}_0\int_0^{t_0}a^{-0.5}(t_e)\int_{t_e}^{t_0} \frac{dt}{a(t)} dt_e,
\end{equation}
\begin{equation}
{\mathit \Phi}_\Lambda(t_0)  =- 4\pi Gc^2 \rho_\Lambda   a_0^{-1.5}\int_0^{t_0}a^{2.5}(t_e)\int_{t_e}^{t_0} \frac{dt}{a(t)} dt_e.
\end{equation}

\subsection{Gravitational potential in Einstein-de Sitter universe}

ES universe is matter dominated, no other energies influence the metric in this model.
Setting the cosmological constant $\Lambda = 0$, one finds that $\mathit{\Phi}_\Lambda=0$ in ES model. Because of the different  scale factors, the age of the universe and  the length parameter are  different in the two models investigated. Quantities indexed by $E$ or $F$  refer to  ES and $\Lambda$CDM (Friedman-Lemaitre) model, respectively.  Denoting the age of the universe $t_{0E}$ as
\begin{equation}
t_{0E}^2= (6\pi G \rho_{m0})^{-1},
\end{equation}
 the scale factor of ES model given  in equation (1.5) takes the form of
\begin{equation}
 a_E(t) =a_{0E}\cdot (t/t_{0E})^{2/3}.
\end{equation}

 Replacing equation (2.10)  into equation (2.5), we obtain  the function of retardation,
\begin{equation}
R_E(t_0,t_e)=\frac{c\, t_{0E}^{2/3}}{a_{0E}} \int_{t_e}^{t_{0E}}\frac{\mathrm{d}t}{t^{2/3}}= \frac{3c\, t_{0E}^{2/3}}{a_{0E}} (t_{0E}^{1/3} - t_e^{1/3})
\end{equation}
in ES model. Substituting equation (2.11) into  equation (2.7)   one gets the gravitational potential of the  observable universe perceived by a fundamental observer,

\begin{equation}\hskip -0.8 cm
{\mathit \Phi}_E(t_0)  =- 12\pi G c^2 \rho_{m0}t_{0E}^{}   \int_0^{t_0E}(t_{0E}^{1/3}t_e^{-1/3}-1) dt_e = -6\pi Gc^2\rho_{m0}t_{0E}^2.
\end{equation}

Replacing the equation (2.9) into (2.12) we obtain the result, the  retarded potential of cosmological gravity in ES universe,  
\begin{equation}
\mathit{\Phi}_E=-\,{c^2}.
\end{equation}
Despite the expansion, the  gravitational potential   does not change on time in ES universe.   In the expanding universe  static gravity is present, which does not depend neither on the observable mass, nor the  extent of the universe. 
Two features of the  potential of cosmological gravity have to be mentioned.   The absolute value of the potential energy of a mass $m$ at rest    relative to the observable universe is equal to the  rest energy, $\vert m\mathit{\Phi}_E\vert =m\,{c^2}$.
On the other hand, the gravitational potential  does not cause gravitational forces  in fundamental frames of reference,  because grad ${\it \Phi} =0$ everywhere, as required by the cosmological principle. A detectable effect of the cosmological gravity  occurs in accelerated frames, as will be shown in section 4.

\subsection{Gravitational potential in $\mathit{\Lambda}$CDM universe}

The relative  matter and vacuum energy density is defined, as
\begin{equation}
\Omega_m = \rho_m/(\rho_m + \rho_\Lambda) \qquad \mathrm{and}\qquad 
 \Omega_\Lambda = \rho_\Lambda/(\rho_m + \rho_\Lambda).
\end{equation}
 $ \Omega_m +  \Omega_\Lambda=1  $, as  it follows from the definition.
    Introducing the symbol $t_\Lambda = 2/\sqrt{3\Lambda}$, used by Gr\o n [12],  the scale factor of $\Lambda$CDM universe  in equation  (1.6) takes the form 
\begin{equation}
a_F(t)= a_{0F}\cdot (\Omega_{m0}/\Omega_{\Lambda0})^{1/3}\cdot  \sinh^{2/3}(t/t_\Lambda).
\end{equation}
The age of $\Lambda$CDM universe can be expressed   by the relative vacuum energy density;

\begin{equation}
 t_{0F}= t_\Lambda \mathrm{artanh}\,  (\sqrt{\Omega_{\Lambda0}}\,),\quad {\rm or}\quad
    t_{0F} = \frac{2}{3H_0}\frac{\mathrm{artanh} \sqrt{\Omega_{\Lambda0}}}{ \sqrt{\Omega_{\Lambda0}}}.
\end{equation}
In present paper the current standard $\Lambda$CDM S model is used with $\Omega_{\Lambda0}=0.728$ and $\Omega_{m 0}=0.272$,  given in  [13]. 

To deduce the gravitational potential in $\Lambda$CDM universe, we  first  determine the function of retardation. Replacing $a_F(t)$ from equation (2.15)  into equation (2.5), the integral of $\mathrm{sinh}^{-2/3}(t/t_\Lambda)$ leads to non-analytical expression, so we use the second order approximation of the  sinus hyperbolic function
\begin{equation}
\mathrm{sinh}(t/t_\Lambda)\approx (t/t_\Lambda)[1+(t/t_\Lambda)^2/6],
\end{equation}
instead. 
This approximation yields the function of retardation as
\begin{eqnarray}
R_F(t_{0F},t_e)= \frac{3c\,t^{2/3}_\Lambda \Omega^{1/3}_{\Lambda0}}{a_{0F}\Omega^{1/3}_{m0}}\Big [t_{0F}^{1/3}-t_e^{1/3}-\frac{1}{63t_\Lambda^{2}}(t_{0F}^{7/3}-t_e^{7/3})\Big].
\end{eqnarray}

\begin{figure}[h]
\begin{center}
\hspace*{10 mm}\includegraphics*[width=120mm]{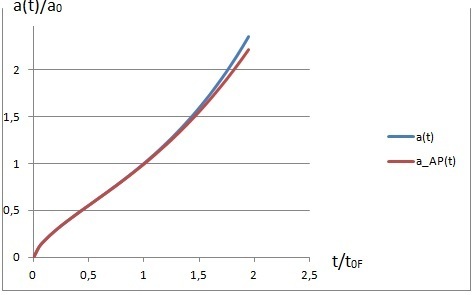}
\vspace*{0 mm}
\center{ \small {\bf Figure 1. } The  scale factor $a(t)$ and its second order approximation, as a function of cosmic time in standard $\Lambda$CDM model. The error of the approximation remains under 1\%, if $t<1.2 t_{0F}$.  }
\end{center}
\end{figure}

Substituting the function of retardation into equations (2.7) and (2.8) and integrating them from zero to present time, it yields the retarded gravitational potentials of matter and vacuum energy. Introducing the abbreviation  $t_{0F}/t_\Lambda=\mathrm{h}(\Omega_{\Lambda0})=\mathrm{artanh}(\sqrt{\Omega_{\Lambda0}})$, they take the form of
\begin{eqnarray}
\hskip -1 cm
 \mathit{\Phi}_{Fm}(t_{0F})=-{c^2}\Big(\frac{\Omega_{m0}}{\Omega_{\Lambda0}}\Big)^{1/2}\mathrm{h}({\Omega_{\Lambda0}})\Big[1-\frac{7}{216}\mathrm{h}^{2}({\Omega_{\Lambda0}})+\frac{7}{3240}\mathrm{h}^4(\Omega_{\Lambda0})\Big], \\
\mathit{\Phi}_{F\Lambda}(t_{0F})=-{c^2}\Big(\frac{\Omega_{m0}}{\Omega_{\Lambda0}}\Big)^{1/2}\mathrm{h}({\Omega_{\Lambda0}})\Big[\frac{1}{12} \mathrm{h}^{2}({\Omega_{\Lambda0}})     +\frac{1}{420}\mathrm{h}^{4}({\Omega_{\Lambda0}})\Big].
\end{eqnarray}

The equation (2.16) holds at any point of cosmic time, so we can express the retarded gravitational potential as a function of relative density of vacuum energy. Summing up the $\mathit{\Phi}_{Fm}(\Omega_{\Lambda})$ and  $\mathit{\Phi}_{F\Lambda}(\Omega_{\Lambda})$         potentials we obtain
\begin{eqnarray}
\hspace{-2.5 cm} \mathit{\Phi}_{F}(\Omega_{\Lambda})=-{c^2}\Big(\frac{1-\Omega_{\Lambda}}{\Omega_{\Lambda}}\Big)^{1/2}
\mathrm{h}(\Omega_{\Lambda})\Big[1+\frac{11}{216}\mathrm{h}^{2}(\Omega_{\Lambda} )+ \frac{1}{216}\mathrm{h}^4 (\Omega_{\Lambda})  \Big] \equiv -\,{c^2}\, f(\Omega_\Lambda). 
\end{eqnarray}

\begin{figure}[h]
\center
\hspace*{0 mm}\includegraphics*[width=120mm]{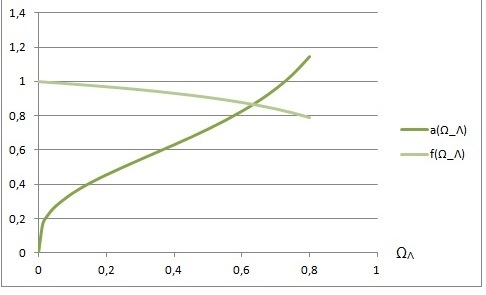}
\begin{center}
\vspace*{0 mm}
\center{ \small {\bf Figure 2. } The plot of the scale factor and the function $f(\Omega_\Lambda)$ versus relative vacuum energy, $ \Omega_\Lambda $ in the standard $\Lambda$CDM  model. The strength, i.e. the absolute value of the gravitational potential decreases,, but its signed value grows as $ \Omega_\Lambda $ increases.  }
\end{center}
\end{figure}

The  model calculation indicates that vacuum energy cannot compensate the fast expansion and cosmological gravity weakens in $\Lambda$CDM   universes.  At the beginning of the universe the gravitational potential starts at $\mathit{\Phi}_F(0)=-\, c^2$, 
and its absolute value weakens monotonically by time, as shown in Figure 2. The present value of the retarded gravitational potential in the standard $\Lambda$CDM universe is 
$ \mathit{\Phi}_{F}(\Omega_{\Lambda 0})=-0.82 c^2 $. This deviation of 18 \% from $-\, c^2$ exceeds  significantly the error of the approximation.

It is worth to note, the growth of the gravitational potential leads to a gravitational redshift of light, $z\approx f(\Omega_{\Lambda e})-f(\Omega_{\Lambda d})$. The subscripts $e$ and $d$ refer to the time of  emission and detection of the photon. 

\section{Cosmological gravity in accelerating frames of reference }

\subsection{Covariant description of cosmological gravitation}

 In the previous section,  strong  potential of cosmological gravity was found in fundamental frames of reference, in ES and $\Lambda$CDM universes.
 Strong gravity is usually associated with significantly curved spacetime.   The presence of a strong cosmological gravity in the locally  Minkowski spacetime perceived by fundamental observers   seems surprising at first. This co-existence  does not conflict with GR, because the spatially  flat FRW metric is generated just by this strong gravity of cosmic matter and   it is considered to be approximately Minkowskian on a local scale only.

 It is obvious, that cosmological gravitation exists not only for  fundamental observers.  Now, we have to find a suitable  description of cosmological gravity, valid even in frames of reference moving relative to the mass distribution of the observable universe.
In Minkowski spacetime SR is valid, even gravitation  obeys it.
Based on this postulate and using SR,  we construct a transformation of cosmological gravitation,  which  ensures   the Lorentz invariance of the motion of a test body in the  cosmological gravitational field, observed from frames of reference moving  relative to fundamental  observers.

\subsection{The Lorentz transformation of the  Lagrangian}

The equivalence of  two different, relative moving, instantaneous inertial frames of reference $K$ and $K'$ from the point of view of the motion of a  test  body  is ensured   by the Lorentz transformation  of the Lagrangian  
\begin {equation}
\gamma_v L = \gamma_{v'} L'.
\end {equation}
  Here $v$ and $ v'$ are the velocities of the test body in $K$ and $K',$ respectively. Hereafter we use the notation $\gamma_v = (1-v^2/c^2)^{-1/2}$ for  arbitrary velocities. 

The validity of equation (3.1) can be easily seen in the cases of a free particle and  of a charged particle moving in electromagnetic field. The first case  is trivial, because  the Lagrangians   of a free test  body of   mass $m$, in two differential frames of reference  $K$ and $K'$ are are given in SR, as
\begin {equation}
L_f= -\frac{mc^2}{\gamma_v}, \qquad L'_f = -\frac{mc^2}{\gamma_{v'}} .
\end{equation}
In the case of a charged particle of mass $m$ and charge of  $q$, the Lagrangians in frames $K$ and $K'$ include the electric scalar and vector potentials, 
\begin {equation}
L_e= -\frac{mc^2}{\gamma_v}-q\mathit{\Phi}_e +q{\bf A_e v}, \qquad L'_e= -\frac{mc^2}{\gamma_{v'}}-q\mathit {\Phi}'_e +q{\bf A_e 'v'}.
\end{equation}
Using SR, one finds the electric potential of a moving, charged source object;  $\mathit{\Phi}_e =\gamma_u \mathit{\Phi}_{Re}$ and  $\mathit{\Phi'}_e =\gamma_{u'} \mathit{\Phi}_{Re}$, where $ \mathit{\Phi}_{Re}$ is  the electric potential in the rest frame of the charged source object.  The vector potentials are  ${\bf A}_e = \mathit{\Phi}_e {\bf u} /c^2$ and ${\bf A}'_e = \mathit{\Phi}'_e {\bf u'} /c^2$, further ${\bf u}$  and ${\bf u'}$  stand for the velocity of the charged source object.
Replacing these quantities into equation (3.3),  using the Lorentz transformation of velocity, an identity is obtained and the validity of equation (3.1) is proved.

\subsection{Lagrangian including cosmological gravitation}

To find a covariant description of cosmological gravity,  the Lagrangian of a test particle  moving in the cosmological gravitational field is required to satisfy   equation (3.1).
Let  the potential of cosmological gravitation be $ \mathit{\Phi}_U   $ in a fundamental frame of reference. In a frame of reference, which is  translating relative to the mass distribution of the observable universe, all the mass elements of the universe move at the same velocity of ${\bf u}$.  The gravitational potential  in a  translating frame of reference   is then $\mathit{\Phi} = \gamma_u^2 \mathit{\Phi}_U$ in  locally Minkowski spacetime.  The coefficient of $ \gamma_u^2$  appears instead of the in electricity usual $ \gamma_u$, because of the mass contribution of the kinetic energy of the source.

Let us construct  the Lagrangian of a test particle of mass $m$ which moves at a velocity of ${\bf v}$ and ${\bf v'}$  in frames $K$ and $K'$. Based on the electromagnetic analogy with equation (3.3), we are looking for the Lagrangian in a form of;
\begin {equation}
\hskip -2.2  cm L_g= -\frac{mc^2}{\gamma_v}-\gamma_v m \mathit{\Phi} + \gamma_v m \frac{\mathit{\Phi}}{c^2}f({\bf u}{\bf v}), \quad L'_g= -\frac{mc^2}{\gamma_{v'}}-\gamma _{v'} m \mathit {\Phi}' +\gamma_{v'} m  \frac{\mathit{\Phi}'}{c^2}f({\bf u'}{\bf v'}).
\end{equation}
Here $\mathit{\Phi} =\gamma^2_u \mathit{\Phi}_{U}$, $\mathit{\Phi'} =\gamma^2_{u'} \mathit{\Phi}_{U}$ and   
 $f$ is an unknown, homogeneous function of the scalar product ${\bf uv}$, it has to  ensure the validity of equation (3.1) in the case of the Lagrangian including cosmological gravitation. 
Let the frame of reference $K$  be  the rest frame of the test particle of mass $m$ and let the frame  $K'$ move at a velocity of ${\bf V}$ relative to $K$. In this case ${\bf v}=0$,  $\gamma_v = 1$ and ${\bf v'} = -{\bf V}$ holds.
 Replacing the quantities $\mathit{\Phi}$ and  $\mathit{\Phi}' $ into equation (3.4) and requiring  $\gamma_vL_g =\gamma_{v'}L'_g$,
\begin {equation}
\gamma^2_u =\gamma^2_{u'}\gamma^2_{v'} -\gamma^2_{u'}\gamma^2_{v'}\frac{f({\bf u'v'})}{c^2}
\end {equation}
is obtained. To eliminate $u $ in equation (3.5),  the Lorentz transformation of velocity is used and  the solution  is  gained, as
\begin {equation}
f({\bf u' v'}) = 2{\bf u' v'} -\frac{({\bf u' v'})^2}{c^2} .
\end {equation}
 Replacing $f$   into equation (3.4),  a covariant description of cosmological gravitation is found. The  relativistic Lagrangian of  a test particle moving in the universe's gravitational field  satisfies the  transformation equation (3.1), if
\begin{equation}
L_g = -\gamma^{-1}_vmc^2 -\gamma^{}_v \gamma^{2}_um\mathit{\Phi}_U\Big(1-\frac{{\bf u v }}{c^2}\Big)^2.
\end{equation}
 Equation (3.7) is  the correct, relativistic Lagrangian of a free test particle, because the cosmological gravitation is even for a free particle present. Up to now, the effect of the cosmological gravitation, i.e. the second term on the RHS in equation (3.7) was simply neglected, and replaced by fictitious inertial forces in accelerated frames of reference.

Excluding relativistic velocities by  setting  $u, \, v \ll c$, expanding equation (3.7) in power series, neglecting the constant terms of $-mc^2$ and $-m\mathit{\Phi}_U$ and terms of order $o(c^{-4})$ we obtain the Newtonian approximation of the Lagrangian,
\begin{equation}
L= m\frac{v^2}{2}-m\mathit{\Phi}_U\frac{v^2}{2c^2}-m\mathit{\Phi}_U\frac{u^2}{c^2}+2m\mathit{\Phi}_U\frac{\bf uv}{c^2}.
\end{equation}
This approximation  holds, even if  the  gravitational potential,  $\mathit{\Phi}_U \sim  c^2$.

The description of cosmological gravitation given by the Lagrangian in equation (3.8) is reminiscent of the weak field, or gravitomagnetic approximation of GR  --  see for example Peacock [14], or  Bini \etal [15] -- but it is not the same. The weak field gravitomagnetism  is based on a linear  approximation  of the metric tensor  in the term of $\mathit\Phi/c^2$,  therefore  that    is applicable for weak gravitation,  but for strong, cosmological gravity  in no case.

The Lagrangian in equation (3.8) differs even from those models, developed by  Bedford and Krumm [16], shortly after by Kolbenstvedt [17] and later by  Karlsson  [18] to    describe the gravitomagnetic interaction between moving masses  using SR  and   Minkowski spacetime.  Above authors neglect the spacetime curvature and  their approach is valid   in weak gravitational fields as well. 

Equation (3.8) is valid  for gravitational interaction between the observable universe and moving test bodies,  in locally Minkowski spacetime.  In the case  of gravitational interaction of masses which forms   Schwarzschild-type, weak inhomogeneities in the FRW universe, only the GR based, weak field approximation gives a correct result -- see
 for example   the gravitomagnetic  interpretation of Gravity Probe B experiment, Vető [19].

\subsection{ Gravitomagnetic approach of cosmological gravity}

Using equation (3.8) we give a gravitomagnetic formulation of cosmological gravity,  introducing scalar and vector potentials. The scalar potential  is already given by  the   potential of cosmological gravitation of $\mathit{\Phi}_U$.

In a frame of reference, which  performs translational motion relative to the mass distribution of the universe at a velocity of $-{\bf u}$,  all the mass element of the observable universe have a uniform instantaneous velocity of  ${\bf u}$  relative to  the frame of reference in question. Because of the uniform velocity of the observable universe in this frame, we can introduce  a gravitational vector potential of 
\begin{equation}
{\bf A}_{T} =\frac{ 2\mathit{\Phi}_U} {c^2}{\bf u},
\end{equation}
 and call it the vector potential of translational motion. 

In the case, when  mass elements of the observable universe move at different velocities relative to the observer, the vector potential of a mass element of the observable universe of  $\mathrm{d}m$, which moves at a velocity of ${\bf u}$ follows from equation (3.9), as
\begin{equation}
\mathrm{d}{\bf A}= \frac{2 \mathrm{d}\mathit{\Phi}}{c^2}{\bf u}, \quad {\rm where}\quad \mathrm{d}\mathit{\Phi}=-G\frac{\mathrm{d} m}{\vert {\bf d'}_F-a_0{\bf R}\vert}
\end{equation}
is the gravitational potential of the mass element.

If the frame of reference  rotates relative to the mass distribution of the universe at a constant angular velocity of ${ {\bomega}}$ and the origin of the rotating frame is located on the axis of rotation, then the  mass elements of the universe at the position vector of ${\bf d}_F$ moves at a velocity of ${\bf u}({\bf d}_F) = {\bf d}_F \times \bomega$ in this frame. 
Replacing  latter in equation (3.10) and summing up the contributions of all the mass elements of the observable universe, the gravitational vector potential in a rotating frame of reference is obtained, as

\begin{equation}
{\bf A}_{R}({\bf R}) = -\frac{2G}{c^2}\int_{V'_{obs}({\bf R})}\frac{{\bf d'}_F\times \bomega\; \rho(t_e)}{\vert{{\bf d'}_F(t_e)-a_0{\bf R}\vert}}dV'.
\end{equation}
In the case of a rotating frame of reference the vector potential depends on the location vector, ${\bf R}$.
Let us transform equation (3.11) into the form of

\begin{equation}\hskip -1.8 cm
{\bf A}_{R}({\bf R}) = -\frac{2G}{c^2}\int_{V'_{obs}({\bf R})}\Big[\frac{({\bf d'}_F - a_0{\bf R})\; \rho(t_e)}{\vert{{\bf d'}_F(t_e)-a_0{\bf R}\vert}}+ \frac{  a_0{\bf R}\; \rho(t_e)}{\vert{{\bf d'}_F(t_e)-a_0{\bf R}\vert}}\Big] dV'\times \bomega.
\end{equation}
The integral of the first term in the square brackets is zero. The mass elements multiplied by a radially directed unit vector are summed up in this integral. The domain of integration is the observable universe centered at the location ${\bf R}$. Because of the spherical symmetry the integral vanishes.
Moving $a_0{\bf R}$ in front of the integral in the second term, the remaining  integral is the scalar potential introduced in equation (2.2). At last, the vector potential of cosmological gravity in a rotating frame of reference is gained, as 
\begin{equation}\hskip -1.8 cm
{\bf A}_{R}({\bf R}) = -\frac{2G}{c^2} a_0{\bf R} \times \bomega \int_{V'_{obs}({\bf R})}\frac{   \rho(t_e)}{\vert{{\bf d'}_F(t_e)-a_0{\bf R}\vert}}dV'=\frac{2  a_0{\bf R} \times \bomega}{c^2}\mathit{\Phi}_U.
\end{equation}

Introducing the notation of  the local vector of position ${\bf r} = a_0{\bf R}$, where  $R\ll R_{max}$,  the vector potential of cosmological gravity in a rotating frame of reference is found,
\begin{equation}
{\bf A}_{R}({\bf r})=\frac{2  \mathit{\Phi}_U}{c^2}  {\bf r} \times \bomega.
\end{equation}
Based on equation (3.14),  the rotation of observable universe seems to be a translation at a velocity of  ${\bf r}\times \bomega$    in rotating frames of reference, at the location of ${\bf r}$.

 Completing equation (3.7) by rotation, we obtain the covariant, relativistic  Lagrangian of a test particle including cosmological gravity
\begin{equation}
L_g = -\gamma^{-1}_vmc^2 -\gamma^{}_v \gamma^{2}_u \gamma^2_{\bf r\times\bomega} m\mathit{\Phi}_U\Big(1-\frac{{\bf u +r\times \bomega }}{c^2}{\bf v}\Big)^2.
\end{equation}
Excluding relativistic velocities, expanding the Lagrangian in power series, as done in the case of equation (3.7) its  Newtonian approximation is obtained. This is the form of a test body's Lagrangian in a translating and rotating frame of reference taking the cosmological gravitomagnetism into account:

\begin{equation} \hskip -1 cm
L= m\frac{v^2}{2}-m\mathit{\Phi}_U\frac{v^2}{2c^2}-m\mathit{\Phi}_U\frac{u^2+({\bf r\times \bomega})^2}{c^2}+2m\mathit{\Phi}_U\frac{\bf (u +r\times \bomega)v}{c^2}.
\end{equation}

\section{Cosmological gravitomagnetism and inertial forces}

\subsection{Inertial forces in Einstein-de Sitter universe}

As given in equation  (2.13) the potential of cosmological gravitation $\mathit{\Phi}_{E}=-c^2$ in ES model universe. Substituting it into equation (3.16), the Lagrangian of a test particle is obtained in a frame of reference, which is translating and rotating relative to the observable universe,

\begin{equation}
L= m{v^2}+m{u^2}+m({\bf r\times \bomega})^2{-2m{\bf (u +r\times \bomega)v}}.
\end{equation}
 To investigate the equation of  motion of the test particle, let us generate the Euler-Lagrange equation, 
\begin{equation}
\frac{{\rm d}}{{\rm d}t}\Big(\frac{\partial L_g}{\partial {\bf v}}\Big)=\frac{\partial L_g}{\partial {\bf r}}.
\end{equation}
Substituting equation (4.1) into (4.2) we obtain
\begin{equation} \hspace{ -1 cm}
\frac{{\rm d}}{{\rm d}t}[2m{\bf v}-2m{\bf (u +r\times \bomega)]}=m\frac{\partial}{\partial {\bf r}}[({\bf r\times \bomega)^2]}-2m\frac{\partial}{\partial {\bf r}}[({\bf r\times \bomega)v]}.
\end{equation}

Getting though with derivation and dividing  by $2m$ in equation (4.3), the acceleration of the test point mass is gained,
\begin{equation}
\dot{\bf v}=\dot{\bf u}-({\bf r\times\bomega})\times \bomega+2{\bf v \times \bomega},
\end{equation}
where the first term is translational, the second one the centrifugal and the third one is the Coriolis acceleration on the RHS of equation (4.4).  The result demonstrates, that inertial forces are caused by the cosmological gravitation in   ES  universes, verifying Mach's suspicions.

\subsection{Inertial forces in $\Lambda$CDM  universe}

The strength of the cosmological  gravity decays with time in $\Lambda$CDM model universe, as it is determined in equation (2.21), as a function of  the relative density of vacuum energy. Replacing  $\mathit{\Phi}_{F}= -c^2  f(\Omega_{\Lambda })$  into equation (3.16), the Lagrangian of a free test body is obtained,
\begin{equation}\hskip -1.2 cm
L=m[1+f(\Omega_\Lambda)]\frac{v^2}{2}+mf(\Omega_\Lambda)[u^2+({\bf r}\times \bomega)^2]- 2mf(\Omega_\Lambda)({\bf u+r \times \bomega)v}.
\end{equation}
Composing the Euler-Lagrange equation and performing the derivations the equation of motion is obtained;
\begin{equation}\hskip -1.2 cm
\dot{\bf v}=\frac{2f(\Omega_\Lambda)}{1+f(\Omega_\Lambda)}[\dot{\bf u}-({\bf r\times\bomega})\times \bomega+2{\bf v \times \bomega}] +\frac{{\rm d} f(\Omega_\Lambda)}{{\rm d}\Omega_\Lambda}\dot{\Omega}_\Lambda(\bf v-u-r\times \bomega).
\end{equation}
In equation (4.6), the  inertial forces  depend on the age of the universe. At present cosmic time  $ f(\Omega_{\Lambda 0})=0.82$ and the coefficient of $2f_0/(1+f_0)=0.901$.
 The model calculation leads to  inertial forces 10 \% weaker than experienced. Based on this result we conclude,  that  $\Lambda$CDM model does not explain the existence of inertial forces correctly.

On the other hand, the  inertial forces in the  second term on the RHS of equation (4.6) are proportional to the velocities ${\bf v, \  u} $ and ${\bf r \times \bomega}$. Therefore, inertial forces are present even in frames of reference performing  rectilinear motion relative to the mass distribution of the universe. These forces occur in consequence of the non-vanishing time derivative of the gravitational potential. The coefficient of $[{{\rm d} f(\Omega_\Lambda)}/{{\rm d}\Omega_\Lambda]}\dot{\Omega}_\Lambda \approx - 6.8\cdot 10^{-19} 1/s$, is so small that it does not cause detectable acceleration in  laboratories.

\section{Results and conclusion}

\subsection{Advantages of retarded cosmological gravitation}

The concept of retarded cosmological gravity introduced in present model seems to be useful in understanding of observed phenomena in cosmology. The mechanism based on it  explains directly how gravitation of the observable  universe gives rise to inertial forces in accelerated frames of reference. The theory presented here, matches  Mach's conception concerning the origin of inertial forces.

In  the  potential field of cosmological gravity each mass in  the universe  has a cosmological potential energy of $ m\mathit{\Phi}_U$.  In ES universe where $\mathit{\Phi}_{E}=-c^2$,   a remarkable coincidence was revealed; the  gravitational potential energy is equal to the rest energy, $ mc^2=-m\mathit{\Phi}_E$. This physical interpretation of the rest energy does not hold exactly in  $\Lambda$CDM universe.

Based on observations, the universe is considered to be locally, i.e.  in a domain of spacetime much smaller than the observable universe, Minkowskian in present model. This realistic approach 
made it possible to find a covariant, relativistic Lagrangian of a free particle moving in the cosmological gravitational field. 
The usual Lagrangian  was  amended by a term (see equations 3.7 and 3.15), which term represents the action of the cosmological gravity on the particle.  In   Newtonian approximation   of the  Lagrangian  a gravitomagnetic model of the cosmological gravitation was given.

\subsection{Inertial forces and Mach's principle in ES and  $\mathit{\Lambda}$CDM universes }

Using the Newtonian approximation of cosmological gravitation, the equation of motion of a test body was investigated. Inertial forces  arose in frames of references accelerated relative to the mass distribution of the observable universe in both of the ES and $\Lambda$CDM models. In ES universe the gravitational potential is constant, $\mathit{\Phi}_{E}=-c^2$ and    inertial forces were find  in agreement with the experimental evidence.

The gravitation is weakening during the expansion   in $\Lambda$CDM model, because the matter content drops faster then  the amount vacuum energy  increases in the observable universe. The potential  starts at $\mathit{\Phi}_F(0)=-\ c^2$ at the big-bang and it runs  to $\mathit{\Phi}_F(t_0)=-0.82 \ c^2$  at present time.
 The strength of the inertial forces depends on the potential of cosmological gravity.  
 Using the present time potential,    the model calculations predicted inertial forces  10 \%  weaker than experimental evidence   in  $\Lambda$CDM universe. This   is a   deficiency  of the  standard model,  because the difference  significantly exceeds the error of the approximation.

 The model calculations indicate that perfect dragging occurs and Mach's principle is valid  in ES model,  but it does not hold   in the present-day standard,  ${\Lambda}$CDM  universe.

\subsection{Comparison with other models}

The results obtained from the model calculations  are in contradiction with Schmid's [8]  statement on the fulfillment of the perfect dragging in all type of  FRW universes.  From the two flat FRW universes investigated in this paper,  perfect dragging   was found  only in the ES model.  Perfect dragging occurs in spatially flat FRW universes  if the potential of cosmological gravitation fulfills the equation  $\mathit{\Phi}_U=-\ c^2$ in present model. This perception is  similar to the Brill-Cohen condition, which gave the criterion  of perfect dragging in Schwarzschild metric, by $\mathit{\Phi}_S=-\ c^2/2$.
Present result seems to confirm Raine's [6] assertion: "Mach's principle holds in FRW universes, in which the metric arises from material source alone".  The vacuum energy indeed cannot  ensure sufficient   gravitational potential   to rise the right  inertial forces, as shown in present calculations.

\Bibliography{1}

\item{ Mach E 1883 {\it Die Mechanik in ihrer Entwicklung} (Leipzig: Brockhaus)}
\item{ Einstein A 1922 {\it The Meaning  of Relativity} (Princeton: University Press)}
\item{  Barbour J and Pfister H (eds) 1995 {\it  Mach’s Principle: From Newton’s Bucket to Quantum Gravity (Einstein
Studies} vol 6) (Boston, MA: Birkhauser) }
\item{ Thirring H 1918, 1921 {\it  Phys. Z.} {\bf 19} 33 and {\bf 22} 29}
\item{  Brill D R and Cohen J M  1966  {\it Phys. Rev. } {\bf 143} 1011-16}
\item{ Raine D J 1975 {\it  Mon. Not. R. astr. Soc.} {\bf 171} 507-528}
\item{ Schmid C 2006 {\it Phys. Rev. } D. {\bf 74} 044031}
\item{  Schmid C 2009 {\it Phys. Rev. } D. {\bf 79} 064007}
\item{ Visser M  and Catto\"en C 2009 {\it Dark Matter in Astrophysics and Particle Physics} Klapdor Kleingrothaus H V and Krivosheina I V (eds) (Published: World Scientific Co) p 287}
\item{ Peacock J A 1998 {\it Cosmological Physics} (Cambridge: University Press) p 67 }
\item{ Berry M 1989 {\it Principles of cosmology and gravitation} (Bristol and Philadelphia: Institute of Physics Publishing) p 123}
\item{ Gr\o n \O \  2008 ({\it Preprint} astro-ph 0801.0552)}
\item{ http://en.wikipedia.org/wiki/Lambda-CDM\_model}
\item{ Peacock J A 1998 {\it Cosmological Physics} (Cambridge: University Press) p 38}
\item{ Bini \etal 2008  {\it Class. Quant. Grav.} {\bf 25} 225014}
\item{Bedford D and Krumm P 1985  {\it Am. J. Phys.} {\bf 53} p 889}
\item{ Kolbenstvedt H 1988 {\it Am. J. Phys.} {\bf 56} p 523}
\item{ Karlsson A 2006 LUTEDX/(TEAT-7150)/1-7 http://www.es.lth.se/teorel/Publications/TEAT-7000-series/TEAT-7150.pdf }
\item{ Vető B 2010 {\it  Eur. J. Phys.} {\bf 31} 1123-30}
\endbib
\end{document}